\documentclass[preprint,12pt]{elsarticle}

\usepackage{amssymb}

\begin{document}

\begin{frontmatter}



\title{Vectorial structure of a hard-edged-diffracted four-petal Gaussian beam in the far field }

\author[label1,label2]{Xuewen Long}
\author{Keqing Lu\corref{cor1}\fnref{label1}}
\cortext[cor1]{Corresponding author. E-mail address:
keqinglu@opt.ac.cn (K.Lu).}
\author[label1,label2]{Yuhong Zhang}
\author[label1,label2]{Jianbang Guo}
\author[label1,label2]{Kehao Li}
\address[label1]{State Key Laboratory of Transient Optics and Photonics, Xi'an Institute of Optics and Precision Mechanics, Chinese Academic of Sciences, Xi'an 710119, China}
\address[label2]{Graduate school of Chinese Academy of Sciences, Beijing, 100039, China}

\begin{abstract}
Based on the vector angular spectrum method and the stationary
phase method and the fact that a circular aperture function can be
expanded into a finite sum of complex Gaussian functions, the
analytical vectorial structure of a four-petal Gaussian beam
(FPGB) diffracted by a circular aperture is derived in the far
field. The energy flux distributions  and the diffraction effect
introduced by the aperture are studied and illustrated
graphically. Moreover, the influence of the $f$-parameter and the
truncation parameter on the non-paraxiality is demonstrated in
detail. In addition, the analytical formulas obtained in this
paper can degenerate into un-apertured case when the truncation
parameter tends to infinity. This work is beneficial to strengthen
the understanding of vectorial properties of the FPGB diffracted
by a circular aperture.
\end{abstract}


\end{frontmatter}


\section{Introduction}
\label{} In recent years, there have been increasing interests in
the study of beam pattern formation. Many beam patterns have
potential and practical applications. For example, many researches
have been done on dark-hollow beam because it is a powerful tool
in precise manipulation and control of microscopic
particles~\cite{Reicherter1999OL,Cai2003OL,Cai2006JOSAA,Wu2008OE,Deng2009JOSAB}.
In fact, many other kinds of special laser patterns, such as
flower-like patterns  and daisy patterns have been observed and
investigated~\cite{Grynberg1994PRL,Berre1996OC}. More recently, a
new form of laser beam called four-petal Gaussian beam has been
introduced and its properties of passing through a paraxial ABCD
optical system have been studied~\cite{Duan2006OC}. Subsequently,
much work has done on four-petal Gaussian
beam~\cite{Gao2006CPL,Chu2008CPL,Tang2008COL,Tang2009JMO,Yang2010OC}.
As is well known, some researches and applications are conducted
in the far field. Meanwhile, in practical applications, however,
there are more or less aperture effects, so it is of practical
significance to study the influence of a hard-edged aperture on
the far-field properties of four-petal Gaussian beam.

In this paper, firstly the power transmissivity of the truncated
four-petal Gaussian beam passing through a circular aperture is
studied. Secondly, the analytical vectorial structure of
four-petal Gaussian beam diffracted by a circular aperture is
derived based on vector angular spectrum
method~\cite{Rosario2001JOSAA,HMGuo2006OE,Zhou2006OL,DengOL},
stationary phase method~\cite{Mandel}, and the fact that a
circular aperture function can be expanded into a finite sum of
complex Gaussian functions~\cite{JJWen}. Based on the analytical
vectorial structure of four-petal Gaussian beam diffracted by a
circular aperture, the energy flux expressions of TE term, TM term
and the whole beam are also obtained, respectively. Thirdly, some
typical numerical examples are given to illustrate the influence
of the  diffraction effect introduced by an aperture on the
far-field energy flux distributions of four-petal Gaussian beam.
Furthermore, the influence of the $f$-parameter and the truncation
parameter on the non-paraxiality is demonstrated. Finally, some
simple conclusions are given.


\section{Analytical vectorial structure of a hard-edged-diffracted four-petal Gaussian beam}
\label{} Let us consider a half space  $z>0$ filled with a linear,
homogeneous, isotropic, and nonconducting medium characterized by
electric permittivity $\varepsilon$ and magnetic permeability
$\mu$. All the sources only lie in the domain  $z<0$. For
convenience of discussion, we consider a four-petal Gaussian beam
with polarization in the x direction, which propagates toward the
half space $z\geq0$ along the z axis. The transverse electric
field distribution of the incident four-petal Gaussian beam at
$z=0$ plane can be written by~\cite{Duan2006OC}

\begin{equation}\label{z0Ex}
E_x(x,y,0)=G_n\left(\frac{xy}{w_0^2}\right)^{2n}\exp\left(-\frac{x^2+y^2}{w_0^2}\right),
n=0,1,2\ldots,
\end{equation}
\begin{equation}\label{z0Ey}
E_y(x,y,0)=0,
\end{equation}
where ${G_n}$ is a amplitude constant associated with the order
$n$; $w_0$ is the
 $1/e^2$ intensity waist radius of the Gaussian
term;  n is the beam order of the four-petal Gaussian beam. When
$n=0$, Eq. (\ref{z0Ex}) reduces to the ordinary fundamental
Gaussian beam with the waist being $w_0$ at the plane $z=0$. The
time factor $\exp(-i\omega t)$ has been omitted in the field
expression. After simple calculation, we know that the distance of
diagonal petals is given by
\begin{equation}\label{distance}
d=2{(2n)}^{1/2}w_0.
\end{equation}
According to Eq.~\ref{distance}, the distance $d$ is determined by
 beam order $n$ and waist size $w_0$.
\begin{figure}[htbp]
\includegraphics[width=14cm]{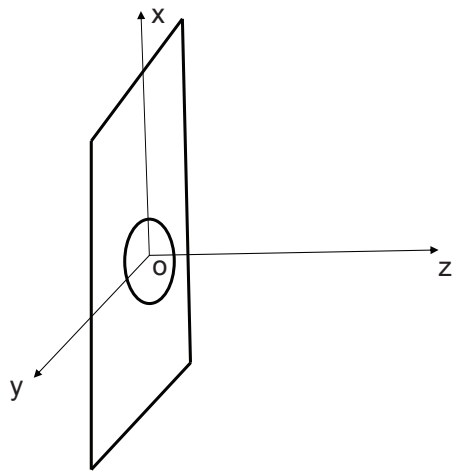}
\caption{Illustrating the geometry of the screen with a circular
aperture and the coordinate system.}\label{fig1}
\end{figure}

Supposing that a hard-edged circular aperture is located at the $z
= 0$ plane, and the center of the circular aperture is the origin.
The geometry for the screen with a circular aperture and the
coordinate system is shown in Fig.~\ref{fig1}. The corresponding
circ function can be written as follows
\begin{equation}
\textmd{circ}(\frac{\rho}{R})=\left\{
\begin{array}{cc}
1,& \rho\leq R,\\
0,& \rho> R.\\
\end{array}
\right.
\end{equation}
where $R$ denotes the radius of the circular aperture and
$\rho=\sqrt{x^2+y^2}$ is radial distance. The power transmissivity
of the truncated four-petal Gaussian beam is given by
\begin{eqnarray}\label{T}
T_n&=&\frac{\int^{R}_{0}\int^{2\pi}_{0}\mid E_x
(\rho,\theta,0)\mid^2\rho d\rho
d\theta}{\int^{\infty}_{0}\int^{2\pi}_{0}\mid E_x
(\rho,\theta,0)\mid^2\rho d\rho d\theta}\nonumber \\
&=&1-\frac{\Gamma(4n+1,2\beta^2)}{4n\Gamma(4n)},
\end{eqnarray}
where $\Gamma(z)=\int^{\infty}_0t^{z-1}e^{-t}dt$  is Euler gamma
function, and $\Gamma(a,z)=\int^{\infty}_zt^{a-1}e^{-t}dt$ is
incomplete gamma function. $\beta=R/w_0$ is defined as the
truncation parameter. When $n$ tends to zero, one can also obtain
the fundamental Gaussian beam power transmissivity, namely
\begin{equation}\label{TT}
T_0=1-\exp(-2R^2/w_0^2).
\end{equation}
\begin{figure}[htbp]
\includegraphics[width=10cm]{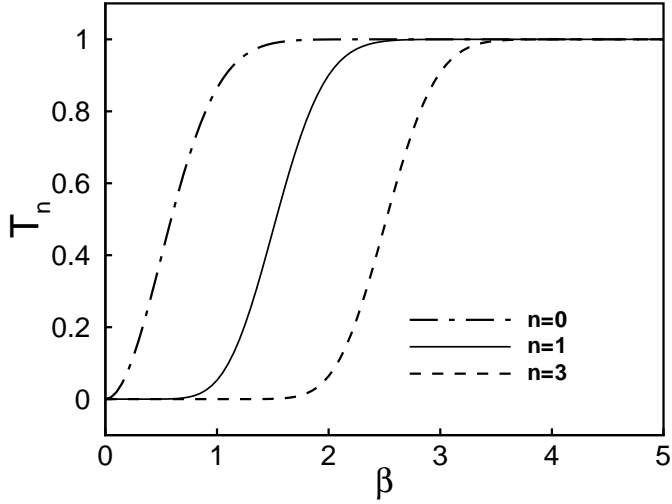}
\caption{The power transmissivity of the truncated FPGB versus the
truncation parameter $\beta$ for various beam order $n =0$
(dash-dotted curve), $n=1$ (solid curve), and $n=3$ (dashed
curve), respectively.}\label{fig2}
\end{figure}
Fig.~\ref{fig2} represents the power transmissivity of the
truncated FPGB versus the truncation parameter $\beta$ for various
beam order $n =0$ (dash-dotted curve), $n=1$ (solid curve), and
$n=3$ (dashed curve), respectively. It is found that $T_n$
increases rapidly with decreasing n for the same value of $\beta$.

Starting from Maxwell's equations, the basic principles of vector
angular spectrum method are vectorial plane wave expansion. The
propagating electric field toward half space $z\geq0$ turns to
be~\cite{Rosario2001JOSAA,DengOL,CGChen2002JOSAA}
\begin{equation}\label{}
E_x(x,y,z)=\int\int^{\infty}_{-\infty}A_x(p,q,\gamma)\exp[ik(px+qy+\gamma
z)]dpdq,
\end{equation}
\begin{equation}\label{}
E_y(x,y,z)=\int\int^{\infty}_{-\infty}A_y(p,q,\gamma)\exp[ik(px+qy+\gamma
z)]dpdq,
\end{equation}
\begin{eqnarray}\label{}
E_z(x,y,z)&=&-\int\int^{\infty}_{-\infty}\left[\frac{p}{\gamma}A_x(p,q,\gamma)+\frac{q}{\gamma}A_y(p,q,\gamma)\right]
 \nonumber \\
&&\times\exp[ik(px+qy+\gamma z)]dpdq,
\end{eqnarray}
where $k=2\pi/\lambda$ denotes the wave number in the medium
related wave length $\lambda$, $\gamma=\sqrt{1-p^2-q^2}$, if
$p^2+q^2\leq1$ or $\gamma=i\sqrt{p^2+q^2-1}$, if $p^2+q^2>1$. The
values of $p^2+q^2<1$ correspond to the homogeneous waves which
propagate at angles $\arcsin\sqrt{p^2+q^2}$ with respect to the
positive z axis, whereas the values of $p^2+q^2>1$ correspond to
the evanescent waves which propagate along the boundary plane but
decays exponentially along the positive z direction. In terms of
Fourier transform, the transverse components of the vectorial
angular spectrum of the electric field just behind the aperture
are expressed as follows

\begin{equation}
A_x(p,q)=\frac{1}{\lambda^2}\int\int^{\infty}_{-\infty}E_x(x,y,0)\textmd{circ}(\frac{\rho}{R})\exp[-ik(px+qy)]dxdy\label{Ax},
\end{equation}

\begin{equation}
A_y(p,q)=\frac{1}{\lambda^2}\int\int^{\infty}_{-\infty}E_y(x,y,0)\textmd{circ}(\frac{\rho}{R})\exp[-ik(px+qy)]dxdy\label{Ay}.
\end{equation}
As is well known, circ function can be expanded into a finite sum
of complex Gaussian
functions~\cite{JJWen,DMZhao2003OC,GQZhou2008OC}
\begin{equation}\label{circ}
\textmd{circ}(\frac{\rho}{R})=\sum_{l=1}^{N}A_l\exp(-B_l\frac{\rho^2}{R^2}),
\end{equation}
where $A_l$, $B_l$ are the coefficients and $N$ is the number  of
complex Gaussian terms; they can be found in Table $1$ of
Ref~\cite{JJWen}. On substituting Eqs. (\ref{z0Ex}), (\ref{z0Ey})
and (\ref{circ}) in Eqs. (\ref{Ax}) and (\ref{Ay}), one can find
\begin{eqnarray}
A_x(p,q)&=&\frac{G_n
}{4\pi^2f^2}\left[\Gamma(n+\frac{1}{2})\right]^2\sum_{l=1}^{N}A_l\left(\frac{\beta^2}{\beta^2+B_l}\right)^{2n+1}
 \nonumber \\
&&\times_1F_1\left(n+\frac{1}{2};\frac{1}{2};-\frac{p^2}{4f^2}\frac{\beta^2}{\beta^2+B_l}\right)
 \nonumber \\
&&\times
_1F_1\left(n+\frac{1}{2};\frac{1}{2};-\frac{q^2}{4f^2}\frac{\beta^2}{\beta^2+B_l}\right)\label{freespace},
\end{eqnarray}
\begin{equation}
A_y(p,q)=0,
\end{equation}
where$f=1/kw_0$, which is $f$-parameter, and $_1F_1(\cdot)$
denotes confluent hypergeometric function. It is well known that
Maxwell's equations can be separated into transverse and
longitudinal field equations and an arbitrary polarized
electromagnetic beam, which is expressed in terms of vector
angular spectrum, is composed of the transverse electric (TE) term
and the transverse magnetic (TM)term, namely,
\begin{eqnarray}
\vec{E}(\vec{r})=\vec{E}_{TE}(\vec{r})+\vec{E}_{TM}(\vec{r})\label{E},
\end{eqnarray}

\begin{eqnarray}
\vec{H}(\vec{r})=\vec{H}_{TE}(\vec{r})+\vec{H}_{TM}(\vec{r})\label{H},
\end{eqnarray}

where
\begin{eqnarray}
\vec{E}_{TE}(\vec{r})&=&\int\int^{\infty}_{-\infty}\frac{1}{p^2+q^2}[qA_x(p,q)-pA_y(p,q)](q\hat{e}_x-p\hat{e}_y)
 \nonumber \\
&&\times\exp(iku)dpdq\label{ETE},
\end{eqnarray}

\begin{eqnarray}
\vec{H}_{TE}(\vec{r})&=&\sqrt{\frac{\varepsilon}{\mu}}\int\int^{\infty}_{-\infty}\frac{1}{p^2+q^2}[qA_x(p,q)-pA_y(p,q)](p\gamma\hat{e}_x+q\gamma\hat{e}_y-b^2\hat{e}_z)
 \nonumber \\
&&\times\exp(iku)dpdq\label{HTE},
\end{eqnarray}
and
\begin{eqnarray}
\vec{E}_{TM}(\vec{r})&=&\int\int^{\infty}_{-\infty}\frac{1}{p^2+q^2}[pA_x(p,q)+qA_y(p,q)](p\hat{e}_x+q\hat{e}_y-\frac{b^2}{\gamma}\hat{e}_z)
 \nonumber \\
&&\times\exp(iku)dpdq\label{ETM},
\end{eqnarray}

\begin{eqnarray}
\vec{H}_{TM}(\vec{r})&=&-\sqrt{\frac{\varepsilon}{\mu}}\int\int^{\infty}_{-\infty}[pA_x(p,q)+qA_y(p,q)]\frac{1}{b^2\gamma}(q\hat{e}_x-p\hat{e}_y)
 \nonumber \\
&&\times\exp(iku)dpdq\label{HTM},
\end{eqnarray}
where $\vec{r}=x\hat{e}_x+y\hat{e}_y+z\hat{e}_z$ is the
displacement vector and $\hat{e}_x$, $\hat{e}_y$, $\hat{e}_z$
denote unit vectors in the x, y, z directions, respectively;
$u=px+qy+\gamma z$; $b^2=p^2+q^2$.

Generally speaking, the evolution of beam is often studied by
virtue of numerical simulation. However, in the far field
framework, the condition $k(x^2+y^2+z^2)^{1/2}\rightarrow\infty$
is satisfied due to z is big enough. By virtue of the method of
stationary phase~\cite{Mandel}, the TE mode and the TM mode of the
electromagnetic field can be given by
\begin{eqnarray}\label{Ete}
\vec{E}_{TE}(\vec{r})&=&-i\frac{G_nZ_R}{\pi}\frac{yz}{r^2\rho^2}\left[\Gamma(n+\frac{1}{2})\right]^2\sum_{l=1}^{N}A_l\left(\frac{\beta^2}{\beta^2+B_l}\right)^{2n+1}
\nonumber
\\&&\times{}_1F_1\left(n+\frac{1}{2};\frac{1}{2};-\frac{x^2}{4f^2r^2}\frac{\beta^2}{\beta^2+B_l}\right)
 \nonumber \\
&&\times{}_1F_1\left(n+\frac{1}{2};\frac{1}{2};-\frac{y^2}{4f^2r^2}\frac{\beta^2}{\beta^2+B_l}\right)
\nonumber \\
&&\times{}\exp(ikr)(y\hat{e}_x-x\hat{e}_y),
\end{eqnarray}

\begin{eqnarray}\label{Hte}
\vec{H}_{TE}(\vec{r})&=&-i\sqrt{\frac{\varepsilon}{\mu}}\frac{G_nZ_R}{\pi}\frac{yz}{r^3\rho^2}\left[\Gamma(n+\frac{1}{2})\right]^2\sum_{l=1}^{N}A_l\left(\frac{\beta^2}{\beta^2+B_l}\right)^{2n+1}
\nonumber \\
&&\times_1F_1\left(n+\frac{1}{2};\frac{1}{2};-\frac{x^2}{4f^2r^2}\frac{\beta^2}{\beta^2+B_l}\right)
 \nonumber \\
&&\times_1F_1\left(n+\frac{1}{2};\frac{1}{2};-\frac{y^2}{4f^2r^2}\frac{\beta^2}{\beta^2+B_l}\right)
 \nonumber \\
&&\times\exp(ikr)(xz\hat{e}_x+yz\hat{e}_y-\rho^2\hat{e}_z),
\end{eqnarray}

and
\begin{eqnarray}\label{Etm}
\vec{E}_{TM}(\vec{r})&=&-i\frac{G_nZ_R}{\pi}\frac{x}{r^2\rho^2}\left[\Gamma(n+\frac{1}{2})\right]^2\sum_{l=1}^{N}A_l\left(\frac{\beta^2}{\beta^2+B_l}\right)^{2n+1}
\nonumber \\
&&\times_1F_1\left(n+\frac{1}{2};\frac{1}{2};-\frac{x^2}{4f^2r^2}\frac{\beta^2}{\beta^2+B_l}\right)
 \nonumber \\
&&\times_1F_1\left(n+\frac{1}{2};\frac{1}{2};-\frac{y^2}{4f^2r^2}\frac{\beta^2}{\beta^2+B_l}\right)
 \nonumber \\
&&\times\exp(ikr)(xz\hat{e}_x+yz\hat{e}_y-\rho^2\hat{e}_z),
\end{eqnarray}

\begin{eqnarray}\label{Htm}
\vec{H}_{TM}(\vec{r})&=&i\sqrt{\frac{\varepsilon}{\mu}}\frac{G_nZ_R}{\pi}\frac{x}{r\rho^2}\left[\Gamma(n+\frac{1}{2})\right]^2\sum_{l=1}^{N}A_l\left(\frac{\beta^2}{\beta^2+B_l}\right)^{2n+1}
\nonumber \\
&&\times_1F_1\left(n+\frac{1}{2};\frac{1}{2};-\frac{x^2}{4f^2r^2}\frac{\beta^2}{\beta^2+B_l}\right)
 \nonumber \\
&&\times_1F_1\left(n+\frac{1}{2};\frac{1}{2};-\frac{y^2}{4f^2r^2}\frac{\beta^2}{\beta^2+B_l}\right)
 \nonumber \\
&&\times\exp(ikr)(y\hat{e}_x-x\hat{e}_y),
\end{eqnarray}
where $r=\sqrt{x^2+y^2+z^2}$, and $Z_R=\pi w_0^2/\lambda$ is
Rayleigh length. Eqs. (\ref{Ete})$-$(\ref{Htm}) are analytical
vectorial expressions for the TE and TM terms in the far field and
constitute the basic results  in this paper. It follows that
spherical wave front remain unchanged for apertured FPGB in the
far field. The results obtained here are applicable for both
non-paraxial case and paraxial case. From Eqs.
(\ref{Ete})$-$(\ref{Htm}), one can find that

\begin{eqnarray}
\vec{E}_{TE}(\vec{r})\cdot\vec{E}_{TM}(\vec{r})=0\label{ETEETM},
\end{eqnarray}
\begin{eqnarray}
\vec{H}_{TE}(\vec{r})\cdot\vec{H}_{TM}(\vec{r})=0\label{HTEHTM}.
\end{eqnarray}
 According to Eqs. (\ref{ETEETM}) and (\ref{HTEHTM}), the TE and TM terms
of a hard-edged-diffracted four-petal Gaussian beam are orthogonal
to each other in the far field.

\section{Energy flux distributions in the far field}
\label{} The energy flux distributions of the TE and TM terms at
the $z=const $ plane are expressed in terms of the z component of
their time-average Poynting vector as
\begin{eqnarray}
\langle
S_z\rangle_{TE}=\frac{1}{2}Re[\vec{E}_{TE}^*\times\vec{H}_{TE}]_z{}\label{SZTE},
\end{eqnarray}
\begin{eqnarray}
\langle
S_z\rangle_{TM}=\frac{1}{2}Re[\vec{E}_{TM}^*\times\vec{H}_{TM}]_z{}\label{SZTM},
\end{eqnarray}
where the $Re$ denotes real part, and the asterisk denotes complex
conjugation. The whole energy flux distribution of the beam is the
sum of the energy flux of  the TE and TM terms, namely
\begin{eqnarray}
\langle S_z\rangle=\langle S_z\rangle_{TE}+\langle
S_z\rangle_{TM}\label{SZ}.
\end{eqnarray}
On substituting Eqs. (\ref{Ete})$-$ (\ref{Htm}) in Eqs.
(\ref{SZTE})$-$ (\ref{SZTM}) yields

\begin{eqnarray}\label{SzTEfluxgenerel}
\langle
S_z\rangle_{TE}=\frac{1}{2}\sqrt{\frac{\varepsilon}{\mu}}\frac{G_n^2Z_R^2}{\pi^2}\frac{y^2z^3}{r^5\rho^2}\left[\Gamma(n+\frac{1}{2})\right]^4\mid
S_n(x,y,z,f,\beta)\mid^2,
\end{eqnarray}

\begin{eqnarray}\label{SzTMfluxgenerel}
\langle
S_z\rangle_{TM}=\frac{1}{2}\sqrt{\frac{\varepsilon}{\mu}}\frac{G_n^2Z_R^2}{\pi^2}\frac{x^2z}{r^3\rho^2}\left[\Gamma(n+\frac{1}{2})\right]^4\mid
S_n(x,y,z,f,\beta)\mid^2.
\end{eqnarray}
Therefore, the whole energy flux distribution of a
hard-edged-diffracted four-petal Gaussian beam in the far field is
given by
\begin{eqnarray}\label{wholeenergyfluxgenerel}
\langle
S_z\rangle=\frac{1}{2}\sqrt{\frac{\varepsilon}{\mu}}\frac{G_n^2Z_R^2}{\pi^2}\frac{z}{r^3\rho^2}\left(\frac{y^2z^2}{r^2}+x^2\right)\left[\Gamma(n+\frac{1}{2})\right]^4
\mid S_n(x,y,z,f,\beta)\mid^2,
\end{eqnarray}
where
\begin{eqnarray}\label{}
S_n(x,y,z,f,\beta)&=&\sum_{l=1}^{N}A_l\left(\frac{\beta^2}{\beta^2+B_l}\right)^{2n+1}
\nonumber \\
 &&\times_1F_1\left(n+\frac{1}{2};\frac{1}{2};-\frac{x^2}{4f^2r^2}\frac{\beta^2}{\beta^2+B_l}\right)
 \nonumber \\
 &&\times_1F_1\left(n+\frac{1}{2};\frac{1}{2};-\frac{y^2}{4f^2r^2}\frac{\beta^2}{\beta^2+B_l}\right),
\end{eqnarray}
the function $S_n(\cdot)$ is defined as above for simplifying
expressions of energy flux.
Eqs.~(\ref{SzTEfluxgenerel})$-$~(\ref{wholeenergyfluxgenerel})
indicate that the diffraction effect introduced by a circular
aperture is described by the truncation parameter $\beta$ in the
far field.  The smaller the truncation parameter is, the more
strongly the field is diffracted by the aperture. In addition, as
the truncation parameter $\beta$ tends to infinity, Eqs.
(\ref{SzTEfluxgenerel}) $-$ (\ref{wholeenergyfluxgenerel})
degenerate into
\begin{eqnarray}\label{SzTEflux}
\langle
S_z\rangle_{TE}&=&\frac{1}{2}\sqrt{\frac{\varepsilon}{\mu}}\frac{G_n^2Z_R^2}{\pi^2}\frac{y^2z^3}{r^5\rho^2}\left[\Gamma(n+\frac{1}{2})\right]^4
{}_1F_1\left(n+\frac{1}{2};\frac{1}{2};-\frac{x^2}{4f^2r^2}\right)^2
 \nonumber \\
&&\times_1F_1\left(n+\frac{1}{2};\frac{1}{2};-\frac{y^2}{4f^2r^2}\right)^2,
\end{eqnarray}

\begin{eqnarray}\label{SzTMeflux}
\langle
S_z\rangle_{TM}&=&\frac{1}{2}\sqrt{\frac{\varepsilon}{\mu}}\frac{G_n^2Z_R^2}{\pi^2}\frac{x^2z}{r^3\rho^2}\left[\Gamma(n+\frac{1}{2})\right]^4
{}_1F_1\left(n+\frac{1}{2};\frac{1}{2};-\frac{x^2}{4f^2r^2}\right)^2
 \nonumber \\
&&\times_1F_1\left(n+\frac{1}{2};\frac{1}{2};-\frac{y^2}{4f^2r^2}\right)^2,
\end{eqnarray}

\begin{eqnarray}\label{wholeenergyflux}
\langle
S_z\rangle&=&\frac{1}{2}\sqrt{\frac{\varepsilon}{\mu}}\frac{G_n^2Z_R^2}{\pi^2}\frac{z}{r^3\rho^2}\left(\frac{y^2z^2}{r^2}+x^2\right)\left[\Gamma(n+\frac{1}{2})\right]^4
\nonumber \\
&&\times{}_1F_1\left(n+\frac{1}{2};\frac{1}{2};-\frac{x^2}{4f^2r^2}\right)^2
 \nonumber \\
&&\times{}_1F_1\left(n+\frac{1}{2};\frac{1}{2};-\frac{y^2}{4f^2r^2}\right)^2.
\end{eqnarray}
As a matter of fact,
Eqs.(\ref{SzTEflux})$-$(\ref{wholeenergyflux}) are energy flux
expressions in un-apertured case, which is not difficult to
understand.
\section{Numerical examples and discussion}
\label{}

\begin{figure}[htbp]
\includegraphics[width=14cm]{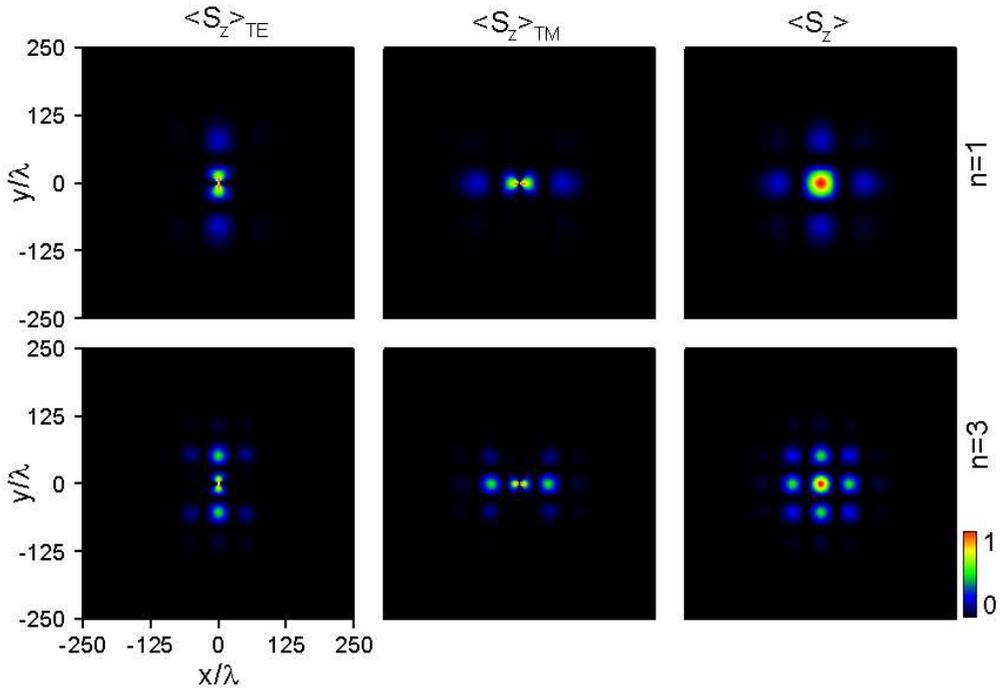}
\caption{(Color online) Normalized energy fluxes $\langle
S_z\rangle_{TE}$, $\langle S_z\rangle_{TM}$ and $\langle
S_z\rangle$ of the FPGB with beam order $n=1$ and $n=3$ (from top
to bottom) in the reference plane $z=2000\lambda$, respectively.
Waist size $w_0$ is set to $10\lambda$, and the circular aperture
radius $R\rightarrow\infty$.}\label{fig3}
\end{figure}

The normalized energy flux distributions of the TE term, the TM
term and  the whole  of the un-apertured beam  at the plane
$z=2000\lambda$ for different beam order $n=1$ and $3$ versus
$x/\lambda$ and $y/\lambda$ are illustrated in Fig.~\ref{fig3}
based on Eqs.~(\ref{SzTEflux})$-$~(\ref{wholeenergyflux}),
respectively. Waist size $w_0$ is set to $10\lambda$ in the
calculation. Apparently, the TE term and the TM term are
orthogonal to each other. The four-petal Gaussian beam splits into
a number of small petals in the far field, which differs from its
initial four-petal shape. Furthermore, the number of petals in the
far field gradually increases when the parameter n increases,
which has potential applications in micro-optics  and beam
splitting techniques, etc~\cite{Duan2006OC}. The above conclusion
is also applicable to  four-petal Gaussian beam diffracted by a
circular aperture. In fact, the four-petal Gaussian beam with beam
order n is not a pure mode, which  can be regarded as a
superposition of $n^2$ two dimensional Hermite-Gaussian
modes~\cite{Duan2006OC}, and different modes evolve differently
within the same propagation distance. The overlap and interference
between different modes result in the propagation properties of
the four-petal Gaussian beam in the far field.

\begin{figure}[htbp]
\includegraphics[width=14cm]{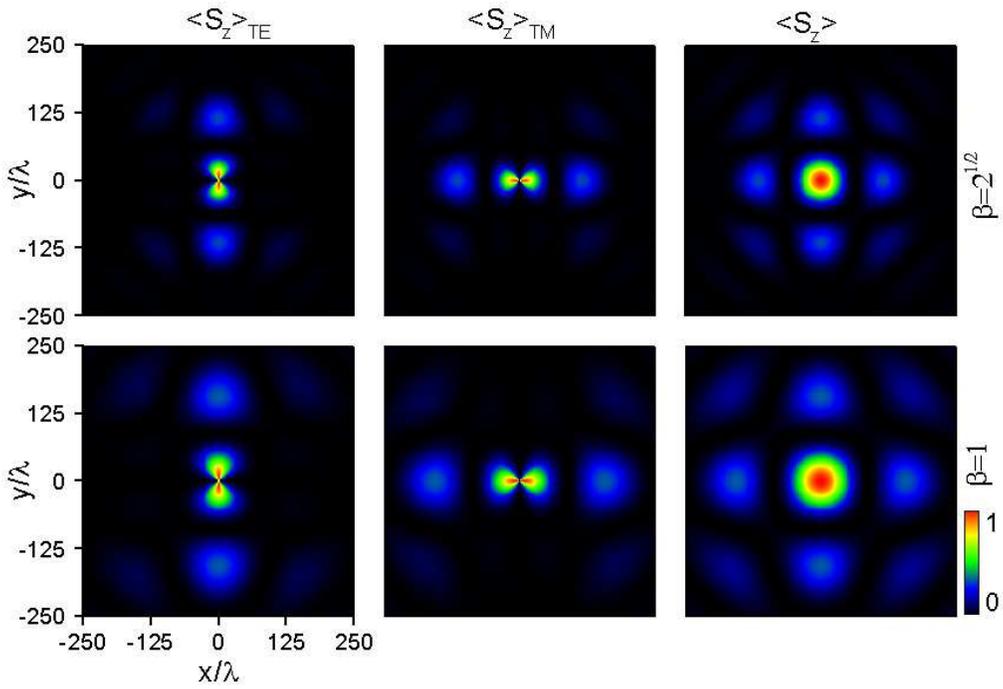}
\caption{(Color online) Normalized energy fluxes $\langle
S_z\rangle_{TE}$, $\langle S_z\rangle_{TM}$ and $\langle
S_z\rangle$ of the hard-edged-diffracted four-petal Gaussian beam
with beam order $n=1$ in the reference plane $z=2000\lambda$. The
truncation parameter is set to $\beta=\sqrt{2}$ and $\beta=1$(from
top to bottom), respectively. Waist size $w_0$ is set to
$10\lambda$.}\label{fig4}
\end{figure}

\begin{figure}[htbp]
\includegraphics[width=14cm]{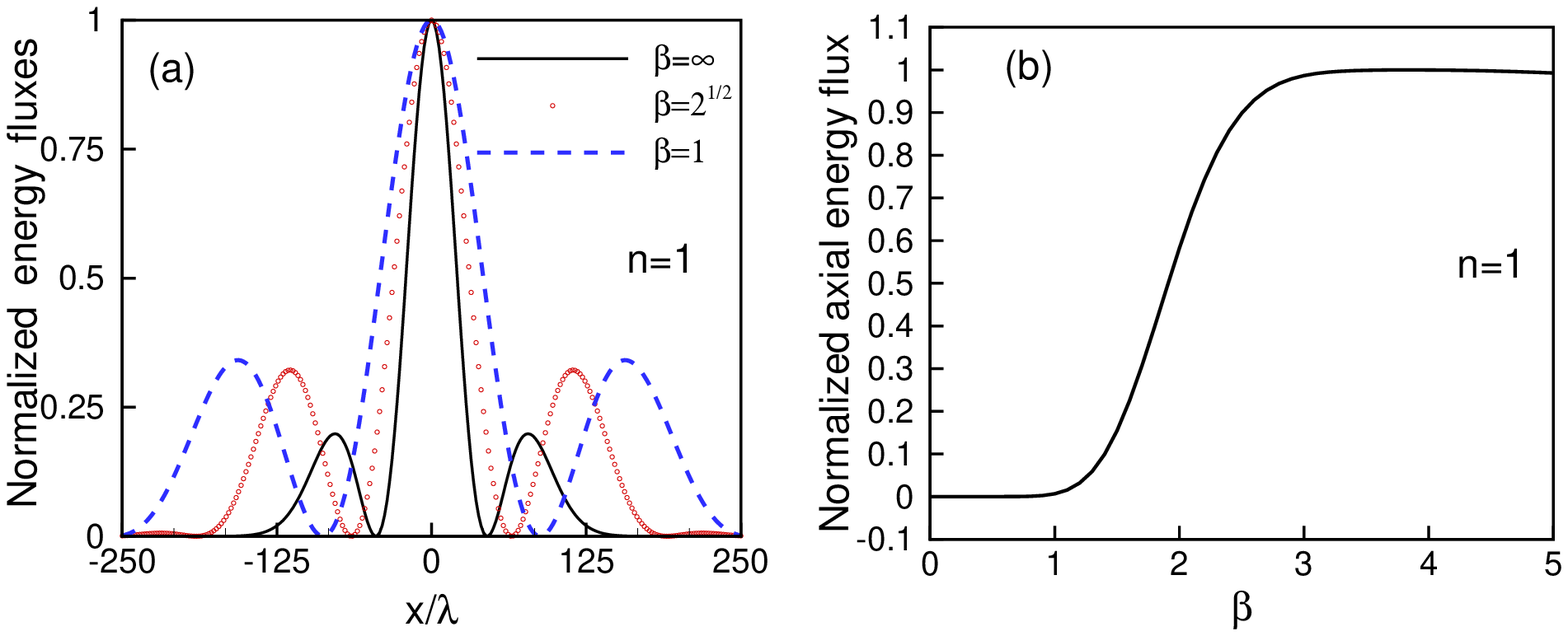}
\caption{(Color online) Cross section and on-axis value of the
normalized total energy flux of the hard-edged-diffracted
four-petal Gaussian beam with the beam order $n=1$ in the
reference plane $z=2000\lambda$. Waist size $w_0$ is set to
$10\lambda$. (a) The cross section with respect to x direction
($y=0$) with various truncation parameter $\beta=1$(dashed curve),
$\beta=\sqrt{2}$(circled curve), and $\beta=\infty$(solid curve),
respectively, (b) Normalized axial energy flux at $z=2000\lambda$
versus the truncation parameter $\beta$.}\label{fig5}
\end{figure}

The influence of the diffraction effect introduced by the aperture
on the energy flux distributions of the four-petal Gaussian beam
is depicted in Fig.~\ref{fig4} and Fig.~\ref{fig5} based on
Eqs.~(\ref{SzTEfluxgenerel})$-$~(\ref{wholeenergyfluxgenerel}).
Waist size $w_0$ is set to $10\lambda$. For simplicity, the beam
order $n$ is set to be $1$. In Fig.~\ref{fig4}, the truncation
parameter is set to $\beta=\sqrt{2}$ and $\beta=1$, respectively.
According to Eq.(\ref{distance}), the four peak-value positions of
the incident beam is just on the boundary line of the circle  when
the truncation parameter is set to $\beta=\sqrt{2}$ for $n=1$.
Comparing top subfigure of Fig.~\ref{fig3} with Fig.~\ref{fig4},
one can find that central spot and side lobes spread more widely
in the far field when the circular aperture exists. Moreover, the
smaller radius of the aperture is, the more wide distribution of
the energy flux is. This phenomenon is easy to understand because
the initial field is confined by the aperture and its diffracted
field has bigger divergence angle in the far
field~\cite{YLZhang2010Optik}. For the sake of showing clearly,
cross section  and on-axis value of the whole energy flux  is
plotted based on Eq.~(\ref{wholeenergyfluxgenerel}) in
Fig.~\ref{fig5}. Fig.~\ref{fig5}(a) shows cross section of the
total energy flux at $y=0$ for different truncation parameter,
from which we can see clearly that the relative values of the
lobes and the full width at half maximum (FWHM) of central spot
become bigger when the truncation parameter decreases. The values
of FWHM are $41.6\lambda$ for $R=\infty$, $61.8\lambda$ for
$\beta=\sqrt{2}$, and $83.8\lambda$ for $\beta=1$, respectively.
In addition, the ratios of the maximum value of the first
side-lobe to that of the central spot are $0.19$ for
$\beta=\infty$, $0.32$ for $\beta=\sqrt{2}$, and $0.34$ for
$\beta=1$, respectively. The on-axis energy flux of the whole beam
at $z=2000\lambda$ versus truncation parameter $\beta$ is shown in
Fig.~\ref{fig5}(b). It should be noticed that truncation parameter
cannot be too small in order to obtain the better transmissivity
in  practical application.
\begin{figure}[htbp]
\includegraphics[width=14cm]{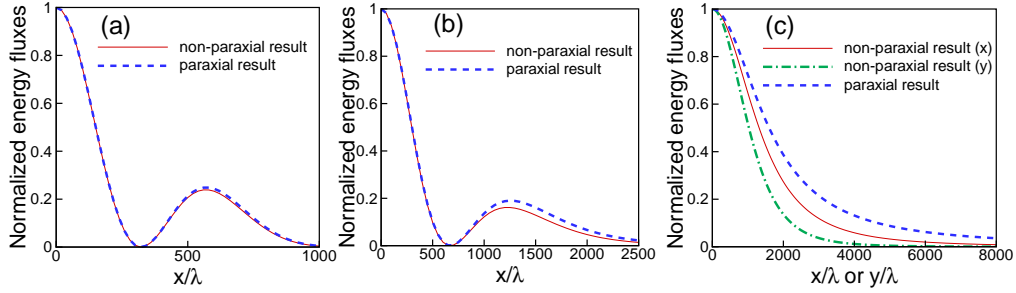}
\caption{(Color online) Cross section at $y=0$ of the normalized
total energy flux of the hard-edged-diffracted four-petal Gaussian
beam with the beam order $n=1$ in the reference plane
$z=2000\lambda$. The solid and the dashed curves denote the
non-paraxial and the paraxial results, respectively. Truncation
parameter $\beta$ is 2. In subfigure(c) cross section at $x=0$ is
also plotted based on non-paraxial result. (a) $f=0.1$, (b)
$f=0.2$, (c) $f=1.5$.}\label{fig6}
\end{figure}

\begin{figure}[htbp]
\includegraphics[width=14cm]{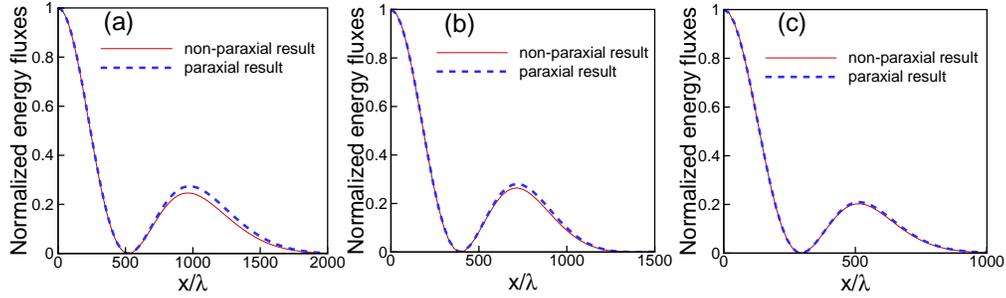}
\caption{(Color online) Cross section at $y=0$ of the normalized
total energy flux of the hard-edged-diffracted four-petal Gaussian
beam with the beam order $n=1$ in the reference plane
$z=2000\lambda$. The solid and the dashed curves denote the
non-paraxial and the paraxial results, respectively.  The
f-parameter is set to be 0.1. (a) $\beta=1.1$, (b) $\beta=1.5$,
(c) $\beta=2.5$.}\label{fig7}
\end{figure}

\begin{figure}[htbp]
\includegraphics[width=14cm]{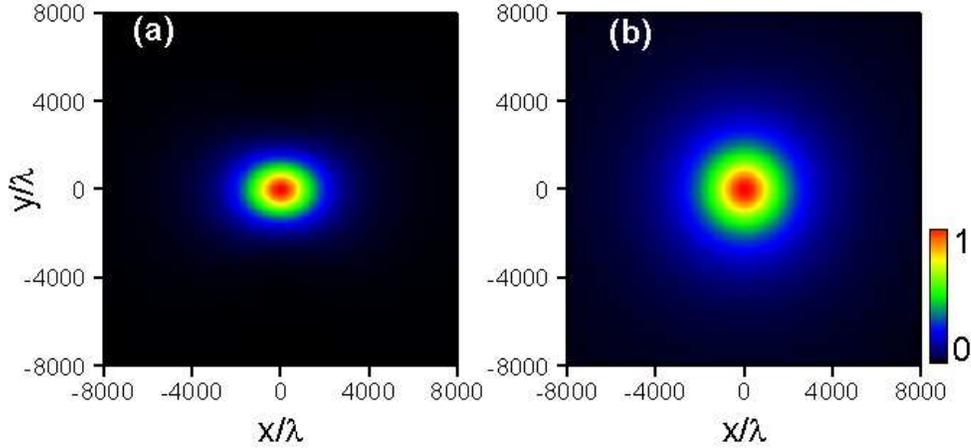}
\caption{(Color online)  Normalized total energy flux of the
hard-edged-diffracted four-petal Gaussian beam with the beam order
$n=1$ in the reference plane $z=2000\lambda$. The $f$-parameter is
set to be 1.5. $\beta=2.0$. (a) non-paraxial result, (b) paraxial
result.}\label{fig8}
\end{figure}

It is well known that the paraxial approximation is allowable when
$w_0/\lambda$ is much larger than 1. The far field expression of
the four-petal Gaussian beam diffracted by a circular aperture
under paraxial regime can be treated as a special case by using
approximation
\begin{eqnarray}\label{paraxialconditon}
r=z+\frac{\rho^2}{2z}\approx z.
\end{eqnarray}
Therefore, the paraxial expression of the whole energy flux of the
four-petal Gaussian beam diffracted by a circular aperture turns
out to be
\begin{eqnarray}\label{paraxialS}
\langle
S_z\rangle=\frac{1}{2}\sqrt{\frac{\varepsilon}{\mu}}\frac{G_n^2Z_R^2}{\pi^2}\frac{1}{r^2}\left[\Gamma(n+\frac{1}{2})\right]^4
\mid S_n(x,y,z,f,\beta)\mid^2.
\end{eqnarray}
Eq.~(\ref{paraxialS}) is symmetric about the x and y variables,
whereas Eq.~(\ref{wholeenergyfluxgenerel}) is somewhat asymmetric
about the x and y variables. With the given beam order, the
non-paraxiality of an apertured FPGB depends on $f$-parameter and
truncation parameter $\beta$. Fig.~\ref{fig6} shows cross section
at $y=0$ of the normalized total energy flux of the
hard-edged-diffracted four-petal Gaussian beam with different
$f$-parameter in the reference plane $z=2000\lambda$. Truncation
parameter $\beta$ is 2. For simplicity, the beam order
$n=1$(hereafter). The solid and the dashed curves denote the
non-paraxial and the paraxial results, respectively(hereafter).
From  Fig.~\ref{fig6}(a) one find that the difference between the
non-paraxial and the paraxial results is negligible for $\beta=2$
and $f\leq1$. The difference between them becomes evident as
$f$-parameter increases.  The central beam spot obtained by
paraxial result is obviously larger than that obtained by
non-paraxial result and side lobes disappear when $f$-parameter
increases enough, which is shown in Fig.~\ref{fig6}(c). Moreover,
the cross section at $x=0$ is also plotted in Fig.~\ref{fig6}(c)
in order to observe the asymmetry. Fig.~\ref{fig7} shows cross
section at $y=0$ of the normalized total energy flux of the
hard-edged-diffracted four-petal Gaussian beam with different
truncation parameter in the reference plane $z=2000\lambda$. The
$f$-parameter is 0.1. The difference between paraxial result and
non-paraxial result becomes evident as truncation parameter
decreases. Comparing Fig.~\ref{fig6} with Fig.~\ref{fig7}, we
obtained a conclusion that the $f$-parameter plays a more key role
in determining the non-paraxiality of an apertured FPGB than does
the truncation parameter $\beta$, which is similar to the
conclusions of Ref.~\cite{Duan2003OE,zhou2010JOSAA}. In order to
understand, Fig.~\ref{fig6}(c) are plotted in contour graphs,
which is shown in Fig.~\ref{fig8}. It can be clearly shown the
non-paraxial result is approximately elliptic, which is obviously
different from the paraxial result. The above conclusion is also
applicable to other higher beam order.

\section{Conclusions}
In summary, the vectorial structure of an apertured four-petal
Gaussian beam in the far field is derived in the analytical form
by using the vector angular spectrum method, the complex Gaussian
expansion of the circular aperture function, and the stationary
phase method. Based on the analytical vectorial structure of an
apertured beam, the energy flux distributions of the TE term, the
TM term and the whole beam are derived in the far-field. Our
formulas obtained in this paper are applicable to both
non-paraxial case and paraxial case. When the truncation parameter
$\beta$ tends to infinity, our formulas degenerate into the
un-apertured case. The four-petal Gaussian beam cannot preserve
its initial shape, and the number of petals in the far field
gradually increases when beam order $n$ increases. Energy
distributions spread more widely in the far field when the
circular aperture exists. The influence of $f$-parameter and
truncation parameter on the far-field behavior is also studied in
detail. The $f$-parameter plays a more key role in determining the
non-paraxiality of an apertured FPGB than does the truncation
parameter $\beta$. In addition, the asymmetry of beam spot becomes
apparently with increasing non-paraxiality. This work is important
to understand the theoretical aspects of vector FPGB and is
beneficial to its practical appliction.
\section*{Acknowledgements}
This research was supported by the National Natural Science
Foundation of China (Grant No.10674176 ).










\end{document}